\documentclass[12pt]{article}

\usepackage{geometry}
\usepackage{amsmath}
\usepackage{amssymb}
\usepackage{amsmath}
\usepackage{amsthm}
\usepackage{bbold}
\usepackage{xspace}
\usepackage{graphicx}

\begin{document}


\thispagestyle{empty}
\begin{flushright}\footnotesize
\texttt{CALT-68-2663}\\
\vspace{0.8cm}
\end{flushright}

\renewcommand{\thefootnote}{\fnsymbol{footnote}}
\setcounter{footnote}{0}

\begin{center}
{\Large\textbf{\mathversion{bold} Sine-Gordon-like action for the Superstring 
in  $AdS_5 \times S^5$ }\par}

\vspace{1.5cm}

\textrm{Andrei Mikhailov and Sakura Sch\"afer-Nameki} \vspace{8mm}

\textit{California Institute of Technology\\
1200 E California Blvd., Pasadena, CA 91125, USA } \\
\texttt{andrei@theory.caltech.edu, ss299@theory.caltech.edu} \vspace{3mm}


\par\vspace{1cm}

\textbf{Abstract}\vspace{5mm}
\end{center}

\noindent
We propose an action for a sine-Gordon-like theory, which reproduces 
the classical equations of motion of the  
Green-Schwarz-Metsaev-Tseytlin superstring on $AdS_5 \times S^5$. 
The action is relativistically invariant. It is a mass-deformed
gauged WZW model for $SO(4,1) \times SO(5) / SO(4) \times SO(4)$
interacting with fermions.

\vspace*{\fill}

\setcounter{page}{1}
\renewcommand{\thefootnote}{\arabic{footnote}}
\setcounter{footnote}{0}


\section{Introduction and Summary}
Quantizing the superstring in $AdS_5 \times S^5$ is important for understanding
string theory in curved spaces and  the AdS/CFT correspondence.
The most successful route so far is to make use of the yet to be proven integrability of
the superstring theory in $AdS_5 \times S^5$ in the light-cone gauge \cite{Metsaev:1998it}. 
But the light-cone gauge-fixed worldsheet theory is a rather unusual theory from the point 
of view of integrable models, as it is not relativistically invariant. Although the 
progress in understanding the worldsheet integrability has so far defied this point, 
it may nevertheless be of interest to obtain a formulation of the theory as a 
relativistically invariant integrable theory. In this paper we will make a step in
this direction.

The idea is to find a reformulation
of the model in terms of a two-dimensional, Lorentz invariant sigma-model, which is
a mass deformation of a conformal field theory. This point of view has been very useful in order to  construct the quantum conserved charges of integrable theories. 
It has been used {\it e.g.} by Reshetikhin and Smirnov \cite{Reshetikhin:1989qg} 
in the context of Sine-Gordon theory and perturbed minimal models, and by
 Bernard and LeClair \cite{Bernard:1990ys} who construct the quantum non-local charges for the sine-Gordon model from the mass-deformed conformal theory of a free boson, or more generally for affine Toda theories by means of mass-deformed WZW models. Key to this approach is that the spectrum of the UV conformal theory is known. Such a formulation of the  Green-Schwarz-Metsaev-Tseytlin (GSMT) string for $AdS_5 \times S^5$ \cite{Metsaev:1998it} is missing, and we wish to propose such a reformulation. We follow the proposal of Bakas, Park and Shin (BPS) \cite{Bakas:1995bm}, which allows to construct for a bosonic symmetric space sine-Gordon model a classically equivalent theory as a mass deformed gauged WZW model.

There are various caveats with this approach, which will require further study.
Firstly this reformulation is on a purely classical level. More precisely, we will construct
a sigma-model, which is similar to the BPS models except that we include fermions.
This sigma-model will reproduce the classical equations of motion of the GSMT superstring
on $AdS_5 \times S^5$. However the Poisson structures of the two theories differ.
Thus, not even classically, these are equivalent theories. 
But surprisingly, this does not yet imply that the quantum theories are different.  
A similar situation occured in \cite{Faddeev:1985qu}, where two different classical 
Poisson structures correspond to expansion around different classical vacua of the same 
quantum model (see also \cite{Smirnov:1982ke}).
Secondly, it would be desirable to obtain a theory that is world-sheet supersymmetric. 
The model that we propose may be world-sheet supersymmetric, however, we were so far 
unable to uncover this structure. It remains to be seen also, whether the perturbation 
of the underlying gauged WZW model can be computed rigorously. 
We leave this for the future.

The plan of this paper is as follows. We first review the boost-invariant symplectic 
structure of the GSMT string (Section \ref{sec:BoostInvariant}). 
Then  we review  the action of Bakas, Park and Shin 
(Sections \ref{sec:BPS} and \ref{sec:InfiniteWorldsheet}) and
discuss subtleties with zero modes and the relation 
to the Hamiltonian reduction of the WZW model (Section \ref{sec:HamiltonianReduction}). 
We then propose (in Section \ref{sec:IntroducingFermions})
 the BPS-type action for the GSMT string in $AdS_5 \times S^5$ and show that 
it reproduces the correct equations of motion. 

\section{The boost-invariant symplectic structure of the Metsaev-Tseytlin
superstring}\label{sec:BoostInvariant}
\subsection{Classical superstring in terms of currents}
The boost-invariant symplectic structure of the classical superstring in
$AdS_5\times S^5$ was constructed in \cite{Mikhailov:2006uc} in the lightcone
formalism. In this formalism the classical string solution is described in terms
of the data on the characteristic. The characteristic is a light-like curve on
the string worldsheet. We will pick a characteristic which is described 
in the conformal 
coordinates $(\tau^+,\tau^-)$ by $\tau^-=0$. With the appropriate choice of the
boundary conditions the string phase space can be described in terms of the
lightcone components of the currents at $\tau^-=0$. The currents $J$ take values
in $\mathfrak{g}=\mathfrak{psu}(2,2|4)$, and the index $0\ldots 3$ indicates the
$\mathbb{Z}_4$ grading 
$\mathfrak{g}=
\mathfrak{g}_0\oplus\mathfrak{g}_1\oplus\mathfrak{g}_{2}\oplus\mathfrak{g}_3$:
\begin{equation}\label{LightConeData}
J_+(\tau^+,0)=
J_{0+} + J_{1+} + J_{2+} + J_{3+} \,.
\end{equation}
There are gauge transformations:
\begin{equation}\label{GaugeTrasformations}
\delta_{\xi}J_+ = \partial_+\xi+[J_+,\xi]\; , \;
\xi\in \mathfrak{g}_0  \,.
\end{equation}
To summarize: 
\begin{eqnarray}
\mathfrak{g}_0 & = & \mathfrak{so}(1,4)\oplus \mathfrak{so}(5) \nonumber \\
\mathfrak{g}_0 + \mathfrak{g}_2 & = & \mathfrak{so}(2,4)\oplus \mathfrak{so}(6) \nonumber\\
\mathfrak{g}_0\oplus\mathfrak{g}_1\oplus\mathfrak{g}_{2}\oplus\mathfrak{g}_3
& = & \mathfrak{psu}(2,2|4) \,.
\nonumber
\end{eqnarray}
We will introduce the notation
\begin{equation}
\nabla_{\pm}=\partial_{\pm} + \mbox{ad}(J_{0\pm}) \,.
\end{equation} 

\subsection{Geometrical meaning of $J_{\pm}$ 
and $\nabla_{\pm}$}\label{GeometricalMeaning}
Geometrically $J_{\bar{2}{\pm}}$ 
are the "lightcone velocity vectors" of the string worldsheet. 
In the near-flat space expansion (see \cite{Mikhailov:2007mr}) they become
 $\partial_{\pm}x+(\vartheta,\Gamma\partial_{\pm}\vartheta)+\ldots$;
both $J_{\bar{2}+}$ and
$J_{\bar{2}-}$ are elements of the tangent space to $AdS_5\times S^5$. 
The $\mathfrak{g}_{\bar{0}}$-components $J_{\bar{0}\pm}$ should be identified
with the Levi-Civita connection (Christoffel symbols). The components in 
$\mathfrak{g}_{\bar{1},\bar{3}}$ are the velocities of the worldsheet fermions, 
they are $J_{\bar{1}\pm}=\partial_{\pm}\vartheta_R+\ldots $ and 
$J_{\bar{3}\pm}=\partial_{\pm}\vartheta_L+\ldots$ in the near-flat space expansion.
(In the flat space limit $\vartheta_L$ and $\vartheta_R$ would come from the
left and right sectors of the worldsheet theory.)

\subsection{Poisson brackets in the lightcone description}\label{PoissonBracketsLightcone}
The $J_+$ components are independent functions of $\tau^+$. The 
$J_-$ components can be, at least formally, expressed through them using
the equations of motion. Therefore the data (\ref{LightConeData}) 
with the gauge equivalence (\ref{GaugeTrasformations}) determines
the string worldsheet. The string worldsheet action is degenerate, and there
are additional local symmetries besides (\ref{GaugeTrasformations}).
The kappa-symmetries are partially fixed by the
conditions
\begin{equation}\label{J1+J3-Zero}
J_{1+}=J_{3-}=0 \,.
\end{equation}
We will assume (\ref{J1+J3-Zero}) throughout this paper. It is useful
to remember that with $J_{1+}=J_{3-}=0$ the equations of motion for $J_2$ are
\begin{equation}
\nabla_{\mp}J_{2\pm}=0 \,.
\end{equation}     
The boost-invariant lightcone Poisson brackets are
\begin{eqnarray}
\{J_{0+},J_{0+}\}^{[0]} & = & 2\nabla_{+}
\\
\{J_{3+},J_{3+}\}^{[0]} & = & -2\mbox{ad}(J_{2+}) \,,
\end{eqnarray}
in the following sense: if $F(J_{0+},J_{3+})$ and $G(J_{0+},J_{3+})$ are two
functionals on the light cone phase space, then their Poisson bracket
is
\begin{equation}
\{ F,G \} = \int d\tau^+\; \mbox{str} \left(
2{\delta F\over \delta J_{0+}} \nabla_{+} {\delta G\over \delta J_{0+}}
-
2{\delta F\over \delta J_{3+}} \left[
J_{2+} , {\delta G\over \delta J_{3+}} \right]
\right) \,,
\end{equation}
with all the other components zero. In particular, the Poisson bracket of $J_{2+}$
with everything else is zero. This means that this Poisson bracket is a degenerate
one, and we have to restrict on the symplectic leaves, see the discussion
in \cite{Mikhailov:2006uc} for details. On a symplectic leaf we have
\begin{equation}
J_{2+}(\tau^+)=J_{2+}^{[0]}(\tau^+) \,,
\end{equation}
where $J_{2+}^{[0]}(\tau^+)$ is a fixed matrix-valued function.
A convenient choice is:
\begin{equation}\label{ChoiceOfJ2Plus}
J_{\bar{2}+}=
\left(\begin{array}{cccccc}
{ 0}	& \alpha_1	&  0	&  0 & 0 & 0 \\
{- \alpha_1}	& 0	& 0	&  0 & 0 & 0 \\
{ 0}	& 0	& 0	&  0 & 0 & 0 \\
{ 0}	& 0	& 0 	&  0 & 0 & 0 \\
{ 0}	& 0	& 0 	&  0 & 0 & 0 \\
{ 0}	& 0	& 0 	&  0 & 0 & 0 
\end{array}\right)_{\mathfrak{so}(2,4)}\!\!\!\bigoplus \;\;	
\left(\begin{array}{cccccc}
{ 0}	& \alpha_2	&  0	&  0 & 0 & 0 \\
{ -\alpha_2}	& 0	& 0	&  0 & 0 & 0 \\
{ 0}	& 0	& 0	&  0 & 0 & 0 \\
{ 0}	& 0	& 0 	&  0 & 0 & 0 \\
{ 0}	& 0	& 0 	&  0 & 0 & 0 \\
{ 0}	& 0	& 0 	&  0 & 0 & 0 
\end{array}\right)_{\mathfrak{so}(6)}	
\end{equation}
where $\alpha_1$ and $\alpha_2$ are some  constants. In string theory we want
$J_{2+}^{[0]}$ to satisfy the Virasoro constraints
\begin{equation}\label{Virasoro}
\mbox{str}\; J_{2+}^2 =0 \,.
\end{equation}
therefore we put:
\[
\alpha_1=\alpha_2\quad (=\mbox{const})\,.
\]
Even after we fix $J_{2+}$ as in (\ref{ChoiceOfJ2Plus}), still $\theta^{[0]}$ is
degenerate. To completely specify the symplectic leaf we fix in addition $J_{3+}$
to be of the form:
\begin{equation}\label{SymplecticLeaf}
J_{3+}-J_{3+}^{(0)}=[J_{2+},K_1] \,,
\end{equation}
with fixed $J_{3+}^{(0)}$.
In the theory of classical superstring in $AdS_5\times S^5$ 
the symplectic leaves of the boost-invariant Poisson bracket are transversal to the 
orbits of the worldsheet reparametrizations and kappa-transformations.
As explained in Section 4.3 of \cite{Mikhailov:2006uc} we can choose the kappa-gauge
so that $J_{3+}^{(0)}=0$, in other words
\begin{equation}\label{J3+SymLeaf}
J_{3+}=[J_{2+},K_1] \,.
\end{equation}
On this symplectic leaf the symplectic form can be written as follows
\begin{equation}\label{SymplecticForm}
\Omega^{[0]}=
\int d\tau^+ \left(\mbox{tr}\; \left(\delta f f^{-1} \delta(\partial_+f f^{-1})
\right) +\mbox{tr}\;\left(\delta K_{\bar{1}} [J_{\bar{2}+},
\delta K_{\bar{1}}]\right)\right) \,.
\end{equation}
where $f$ is related to $J_{0+}$ by the formula
\begin{equation}
J_{0+}=-\partial_+ f f^{-1} \,.
\end{equation}
The discussion in \cite{Mikhailov:2006uc} was limited to the positive component
of the lightcone: $\tau^-=0$. To completely describe the worldsheet, we have to specify
a second characteristic, that is the negative component of the lightcone: $\tau^+=0$.
We can choose 
\begin{equation}\label{J1-SymLeaf}
J_{1-}=[J_{2-},K_3] \,,
\end{equation}
on the negative component. Then the equations of motion
are compatible with (\ref{J3+SymLeaf}) and (\ref{J1-SymLeaf}) in the following way
\begin{eqnarray}
\nabla_-K_1&=&J_{1-}+X_{1-}  \\
\nabla_+K_3&=&J_{3+}+X_{3+}\,,
\end{eqnarray}
where $X_{1-}$ and $X_{3+}$ are undetermined quantities with the property
$[J_{2+},X_{1-}]=[J_{2-},X_{3+}]=0$.

\section{The action giving rise to the boost-invariant Poisson bracket}
\label{sec:Action}
In this section we rewrite the classical string equations of motion 
in a form which closely resembles the equations of motion of a
gauged WZW model with a mass term. Then we will
explain what is precisely the relation.

\subsection{The action of Bakas, Park and Shin}
\label{sec:BPS}
\subsubsection{An equivalent form of the string worldsheet equations of motion.}
As a warmup let us consider the bosonic 
string on $\mathbb{R}\times S^n$. 
The sphere $S^n$ is the symmetric space 
$SO(n+1)/SO(n)$. We denote:
\begin{equation}
 G=SO(n+1)\quad, \quad G_0=SO(n)\quad , 
\quad H=SO(n-1) \,.
\end{equation}
The corresponding Lie algebras are:
\begin{eqnarray}
\mathfrak{g}=\mathfrak{g}_2\oplus \mathfrak{g}_0 & = & \mathfrak{so}(n+1) \nonumber
\\
\mathfrak{g}_0 & = & \mathfrak{so}(n) \nonumber
\\
\mathfrak{h} & = & \mathfrak{so}(n-1) \,.\nonumber
\end{eqnarray}
The equations of motion are
\begin{eqnarray}
\nabla_+J_{2-}=\nabla_-J_{2+}&=&0
\label{EqM2}
\\[1pt]
[\nabla_+,\nabla_-]+[J_{2+},J_{2-}] &=& 0 \,,
\label{EqM0}
\end{eqnarray}
where $\nabla_{\pm}=\partial_{\pm}+J_{0\pm}$.
We can choose such a gauge that $J_{2+}=T$ is a constant matrix 
({\it cf.} Eq. (\ref{ChoiceOfJ2Plus})). For example for $n=5$ we can take:
\[
T=\left(\begin{array}{cccccc}
{ 0}	& 1	&  0	&  0 & 0 & 0 \\
-1	& 0	& 0	&  0 & 0 & 0 \\
{ 0}	& 0	& 0	&  0 & 0 & 0 \\
{ 0}	& 0	& 0 	&  0 & 0 & 0 \\
{ 0}	& 0	& 0 	&  0 & 0 & 0 \\
{ 0}	& 0	& 0 	&  0 & 0 & 0 
\end{array}\right)
\]
Then 
the stabilizer of $T$ in $\mathfrak{g}_0$ is $\mathfrak{h}=\mathfrak{so}(n-1)$.
Then (\ref{EqM2}) implies that 
\begin{equation}
J_{0-}=A_-\in \mathfrak{h} \,.
\end{equation}
Let us introduce $g\in G_0$ such that 
\begin{equation}\label{J2-VsG}
J_{2-}=g^{-1}J_{2+}g \,.
\end{equation}
Then Eq. (\ref{EqM2}) implies that
\begin{equation}
\partial_+ + J_{0+}=g^{-1}(\partial_+ + A_+)g\;,\;
A_+\in \mathfrak{h} \,.
\end{equation}

\vspace{20pt}
\noindent
Therefore the phase space of the classical string can be described
by the data
\begin{equation}\label{gAAdata}
(g\;,\;A_+\;,\; A_-) \,,
\end{equation}
subject to the equations
\begin{equation}\label{MassiveZeroCurvature}
[g^{-1}(\partial_++A_+)g\;,\;\partial_-+A_-]+
[T,g^{-1}Tg]=0 \,,
\end{equation}
modulo the gauge symmetries
\begin{eqnarray}
 g&\mapsto& h_L g h_R^{-1}
\nonumber
\\
 \partial_+ + A_+&\mapsto& h_L (\partial_+ + A_+) h_L^{-1}
\label{LeftRightGaugeTransformations}
\\
 \partial_- + A_-&\mapsto& h_R (\partial_- + A_-) h_R^{-1} \,.
\nonumber
\end{eqnarray}
The $h_L$ gauge symmetry is a "tautological" gauge symmetry, existing because
we replaced $J_{2-}$ with $g$, see Eq. (\ref{J2-VsG}). And the $h_R$ gauge
symmetry is what remains of (\ref{GaugeTrasformations}), after we put $J_{2+}=T$.

\subsubsection{Gauged WZW with a mass term}
\label{sec:GWZW}
Eq. (\ref{MassiveZeroCurvature}) is identified in \cite{Bakas:1995bm}
as one of the equations of motion of the mass deformed gauged WZW model with the gauge
fields $A_+$ and $A_-$. 
More precisely, the action takes the form
\begin{equation}\label{ActionBPS}
S_{BPS} (g, A_+, A_-) = S_{WZW}(g) + S_{gauge} (g, A_+, A_- )
+ S_{mass}(g)\,,
\end{equation}
where
\begin{equation}\label{TermsInWZWAction}
\begin{aligned}
S_{WZW}   &= -{1\over 4\pi} \left(
              \int  d^2\tau \hbox{Tr} (\partial_+ g \partial_- g^{-1})
              + \int_B {1\over 3} \hbox{Tr} (g^{-1} d g)^3  \right)\cr
S_{gauge} &=  {1\over 2 \pi} 
              \int d^2\tau   \hbox{Tr} \left(
              A_+A_- - A_+gA_-g^{-1} + A_+\partial_-gg^{-1} -A_-g^{-1}\partial_+g) \right)\cr
S_{mass}  &=  {1\over 2\pi}
              \int d^2\tau   \hbox{Tr} \left( Tg^{-1}Tg \right) \,.
\end{aligned}
\end{equation}
The variation with respect to $g$ of the action $S_{BPS}(g,A_+,A_-)$ is
\begin{equation}
\delta S_{BPS} = 
\int \hbox{Tr}\;
\left( 
([g^{-1}(\partial_++A_+)g\;,\;\partial_-+A_-]+
[T,g^{-1}Tg])  g^{-1}\delta g
\right)  \,.
\end{equation}
This leads to the equation of motion which is identical to 
(\ref{MassiveZeroCurvature}):
\begin{equation}\label{GWZWEqM}
[g^{-1}(\partial_++A_+)g\;,\;\partial_-+A_-]+
[T,g^{-1}Tg]=0 \,,
\end{equation}
The variation with respect to $A_+$ and $A_-$ gives the equations of motion
for the gauge fields
\begin{eqnarray}
A_+&=&\left(g^{-1}(\partial_++A_+)g\right)_{\mathfrak{h}}\label{AplusEqM}
\\
A_-&=&\left(g(\partial_-+A_-)g^{-1}\right)_{\mathfrak{h}} \,.\label{AminusEqM}
\end{eqnarray}

We will explain the relation between the classical string described
by the equations
of motion (\ref{gAAdata}) --- (\ref{LeftRightGaugeTransformations})
and the gauged WZW model. The main idea is to observe that the classical
string can be identified with the Hamiltonian reduction of the 
WZW model with respect to the symmetries (\ref{SymmetriesToReducePrime}).
The Hamiltonian reduction of the WZW model is closely related
to the gauged WZW model, in fact it coincides with the
gauged WZW model up to subtleties with zero modes. 

\subsection{Relation between string worldsheet theory and gauged WZW: formal analysis
on an infinite worldsheet}
\label{sec:InfiniteWorldsheet}
Classical solutions of the action (\ref{ActionBPS}) are also solutions  
of Eqs. (\ref{MassiveZeroCurvature}). It is not immediately
obvious why all the solutions of (\ref{gAAdata}) --- (\ref{LeftRightGaugeTransformations}) 
can be obtained as classical solutions of (\ref{ActionBPS}), because there are additional 
equations (\ref{AplusEqM}) and (\ref{AminusEqM}). In other words, the action
(\ref{ActionBPS}) gives solutions of the system 
(\ref{gAAdata}) --- (\ref{LeftRightGaugeTransformations}) with the particular
$A_{\pm}$, namely $A_{\pm}$ satisfying (\ref{AplusEqM}) and (\ref{AminusEqM}).
One has to prove that any solution of (\ref{MassiveZeroCurvature})
can be transformed by the gauge transformations (\ref{LeftRightGaugeTransformations})
to a solution satisfying (\ref{AplusEqM}) and (\ref{AminusEqM}).
The detailed analysis of this question in both bosonic
and supersymmetric cases is discussed in \cite{GrigorievTseytlinForthcoming}.
Let us briefly summarize the argument from our point of view.
We have to prove that all the solutions of the system
(\ref{gAAdata}) --- (\ref{LeftRightGaugeTransformations}) can be obtained
from (\ref{ActionBPS}).
Given an arbitrary solution of the system 
(\ref{gAAdata}) --- (\ref{LeftRightGaugeTransformations}), we can bring it
to the gauge $A_{\pm}=0$ using the gauge transformations 
(\ref{LeftRightGaugeTransformations}) with the parameters
$h_L=P\exp\int d\tau^+A_+$ and $h_R=P\exp\int d\tau^-A_-$. 
In this gauge $g$ satisfies
\begin{equation}\label{EqMGaugeA0}
\partial_-(g^{-1}\partial_+g)=[T,g^{-1}Tg] \,.
\end{equation}
Moreover, even after we fix the gauge $A_{\pm}=0$ there are still residual
gauge transformations with $h_L=h_L(\tau^-)$ and $h_R=h_R(\tau^+)$.
Let us first assume that the worldsheet is infinite, then
we can use these gauge transformations to further fix the gauge, 
so that
\begin{eqnarray}
(g^{-1}\partial_+ g)_{\mathfrak{h}} &=&0
\label{HolCurrentMustVanish}
\\
(\partial_-g g^{-1})_{\mathfrak{h}} &=&0 \,.
\label{AntiHolCurrentMustVanish}
\end{eqnarray}
This is possible because (\ref{EqMGaugeA0}) implies that
\begin{equation}\label{DefinitionOfJPlusAndJMinus}
j_+=(g^{-1}\partial_+ g)_{\mathfrak{h}}\;\; \mbox{and} \;\;
  j_-=(\partial_-g g^{-1})_{\mathfrak{h}} \,,
\end{equation}
are holomorphic and antiholomorphic currents
\begin{equation}\label{ConservationOfCurrents}
\partial_-j_+ =\partial_+j_- = 0 \,.
\end{equation}
This means that $h_R=P\exp \left(-\int j_+d\tau^+\right)$ is holomorphic
and $h_L=P\exp\left(-\int j_-d\tau^-\right)$ is antiholomorphic and therefore
we can use the residual gauge transformation with these $h_L$ and $h_R$ to
fix $j_+=j_-=0$ which is precisely 
(\ref{HolCurrentMustVanish}), (\ref{AntiHolCurrentMustVanish}).
Now $(g,A_{\pm})$ satisfies (\ref{AplusEqM}) and (\ref{AminusEqM}).
This proves that any solution of the classical string
equations (\ref{gAAdata}) --- (\ref{LeftRightGaugeTransformations})
can be gauge transformed to satisfy (\ref{AplusEqM}) and (\ref{AminusEqM}),
and therefore is also a solution of the equations of motion
of (\ref{ActionBPS}).

\subsection{Interpretation as Hamiltonian reduction of WZW model}

In the next section we will include fermions and explain how the classical
superstring in $AdS_5\times S^5$ is related to the gauged WZW model interacting
with fermions. But before we discuss fermions we want to give
a ``geometrical'' explanation of why Eq. (\ref{EqMGaugeA0}) implies 
the existance of the holomorphic and antiholomorphic
currents, Eq. (\ref{ConservationOfCurrents}).
This will be useful for understanding the fermionic extension.
After we include fermions the holomorphic and antiholomorphic
currents become more complicated, but the geometrical interpretation of them as
moment maps remains the same.

Notice that Eq. (\ref{EqMGaugeA0}) is the equation of motion of the mass deformed
(ungauged) WZW model with the action $S_{WZW}(g)+S_{mass}(g)$. 
The classical phase space of the WZW model (with or without
the mass term) has a symmetry:
\begin{equation}\label{SymmetriesToReduceInMainText}
g(\tau^+,\tau^-)\mapsto h_L(\tau^-) g(\tau^+,\tau^-) h_R(\tau^+)^{-1} \,.
\end{equation}
Here $h_L(\tau^-)$ and $h_R(\tau^+)$ are periodic $H$-valued functions,
so the symmetry group is the product of two loop groups: $LH\times LH$.
Now we have a Hamiltonian system (the classical WZW model) and a symmetry
acting on its phase space ($LH\times LH$). One can verify that this symmetry
preserves the symplectic form, and in fact the currents $j_+$ and $j_-$ defined
by Eqs. (\ref{DefinitionOfJPlusAndJMinus})  are precisely the densities
of the moment map corresponding to this symmetry, and Eq. (\ref{ConservationOfCurrents})
is the conservation of the moment map. (See the Appendix for details.)

Setting $j_{\pm}$ to zero (Eqs. (\ref{HolCurrentMustVanish}) and 
(\ref{AntiHolCurrentMustVanish}))
corresponds to considering the {\em Hamiltonian reduction} of the WZW model
by the symmetry $LH\times LH$.
From this point of view the classical string described by Eqs. 
(\ref{gAAdata}) --- (\ref{LeftRightGaugeTransformations})
is naturally identified, at least at the level of equations of motion,
with the Hamiltonian reduction of the WZW model.
The Hamiltonian reduction of the WZW model is closely related to the gauged
WZW model (\ref{ActionBPS}), in fact it is equivalent to the gauged WZW model 
on the infinite
line. 
On the cylinder there is a mismatch of zero modes, see the Appendix for details.

\vspace{20pt}
\noindent
Similar arguments hold for the fermionic extension which we will now describe.

\section{Including fermions}
\label{sec:IntroducingFermions}

We will now show that we can include fermions to the mass-deformed gauged WZW model so that the classical equations of motion agree with those of the Metsaev-Tseytlin string in $AdS_5 \times S^5$.

\subsection{Fermionic terms in the action}

The symplectic leaf (\ref{SymplecticLeaf}) 
is parametrized by $K_1$.
This means that the action of the generalized sine-Gordon model should be
described in terms of $K_{1}$ and $K_{3}$ rather  than
$J_{3}$ and $J_{1}$. Therefore we will take $K_{1}$ and $K_{3}$
as independent variables.
Eq. (\ref{SymplecticForm}) suggests to look for an action in the following 
form:
\begin{eqnarray}\label{ActionSSSG}
S&=&S_{BPS}+\Delta S_{kin}+\Delta S_{mass}
= S_{BPS}+\int d^2\tau \left\{ 
-{1\over 2}\mbox{str} [J_{2+},K_{1}]\nabla_-K_{1} - \right.\nonumber\\
&&\left. -{1\over 2}\mbox{str} [J_{2-},K_{3}]\nabla_+K_{3} 
+\mbox{str}[J_{2+},K_{1}][J_{2-},K_{3}]\right\} \,,
\end{eqnarray}
where $S_{BPS}$ is described in the previous section.
Let us consider the gauge where $J_{2+}=T$ and $J_{2-}=g^{-1}Tg$
and $T$ is a constant matrix. In this gauge 
\begin{eqnarray}
&&	\nabla_-=\partial_-+\mbox{ad}(J_{0-})
=\partial_-+\mbox{ad}(A_-) \\
&&	\nabla_+=\partial_++\mbox{ad}(J_{0+})=
\partial_++\mbox{ad}(g^{-1}A_+g +g^{-1}\partial_+ g) \,.
\end{eqnarray}
We will now prove that this action leads to the correct equations of motion for the
classical superstring in $AdS_5\times S^5$.

\subsection{Equations of motion}

We now derive the equations of motion from the variation of $K_{1,3}$ and $g$ and show that these agree with the string equations of motion. 

First consider the variation with respect to the fields $K_{1}$ and $K_3$, which will yield 
the equations of motion for the fermionic fields.
Varying  $\delta K_1$ gives
\begin{equation}\label{J3EOM}
-[J_{2+},\nabla_-K_1]+[J_{2+},[J_{2-},K_3]]=
-\nabla_-J_{3+}-[J_{1-},J_{2+}]=0 \,.
\end{equation}
Likewise the variation with respect to $K_3$ yields
\begin{equation}\label{J1EOM}
-[J_{2-}, \nabla_+ K_3] + [J_{2-}, [J_{2+}, K_1]]
= - \nabla_+ J_{1-} - [J_{3+}, J_{2-}] =0 \,.
\end{equation}
These are the correct equations of motion for the fermions.

The bosonic equations are obtained from the variation $\delta_{\xi}g=g\xi$. We have
\begin{equation}
\delta_{\xi}\nabla_-=0, \;\;\;
\delta_{\xi}\nabla_+=\mbox{ad}(\nabla_+\xi),\;\;\;
\delta_{\xi}J_{2-}=[J_{2-},\xi]  \,.
\end{equation}
The $\xi$-variation of the BPS action gives
\begin{equation}
\delta_{\xi}S_{BPS}=
\int \mbox{str}\left(\xi\left( \partial_+ J_{0-} -\partial_- J_{0+}
+[J_{0+},J_{0-}]+[J_{2+},J_{2-}]\right)\right) \,.
\end{equation}
The variation of $\Delta S_{kin}$ is
\begin{eqnarray}
\delta_{\xi}\Delta S_{kin} &=& -{1\over 2}\int \mbox{str}
\xi \nabla_+[K_3,[K_3,J_{2-}]]
+{1\over 2}\int \mbox{str}\xi [J_{2-},[K_3,\nabla_+K_3]]=\nonumber\\
&=& -\int \mbox{str} \xi [K_3,[\nabla_+K_3,J_{2-}]]
=\int\mbox{str} \xi [K_3,[J_{2-},J_{3+}]] \,.
\end{eqnarray}
We used the fermion equation of motion (\ref{J1EOM}), 
which implies that $\nabla_+ K_3 = J_{3+} +$ terms that are annihilated by ad$(J_{2-})$.
Finally, the variation of $\Delta S_{mass}$ is
\begin{equation}
\delta_{\xi}\Delta S_{mass}=
\int\delta_{\xi}\; \mbox{str} [J_{2+},K_1][J_{2-},K_3] =
- \int\mbox{str} [ [ [J_{2+},K_1], K_3],J_{2-}]\xi \,.
\end{equation}
We get:
\begin{equation}
\delta_{\xi}(\Delta S_{kin} + \Delta S_{mass})=
\mbox{str} \left(\xi [J_{3+},J_{1-}]\right) \,.
\end{equation}
Therefore the variation of $S_{BPS}+\Delta S_{kin} +\Delta S_{mass}$
 gives the correct bosonic equation of motion
\begin{equation}\label{J2EOM}
\partial_+ J_{0-}-\partial_- J_{0+}+[J_{0+},J_{0-}]+[J_{2+},J_{2-}]
+[J_{3+},J_{1-}]=0 \,.
\end{equation}
We have shown that the action  of (\ref{ActionSSSG}) reproduces 
correctly all the equations of motion, (\ref{J1EOM}), (\ref{J3EOM}) 
and (\ref{J2EOM}), of the GSMT super-string on $AdS_5 \times S^5$. 

\subsection{Variation with respect to $A_{\pm}$}

The story is similar to the case of the bosonic string. 
As in Section \ref{sec:GWZW} we can go to the gauge where $A_{\pm}=0$.
The equations following from (\ref{ActionSSSG}) are the same as the 
string equations of motion plus the vanishing of the holomorphic current
\begin{equation}
j_+=\left(g^{-1}\partial_+g - {1\over 2} [K_1,[K_1,J_{2+}]]\right)_{\mathfrak{h}} \,,
\end{equation}
and vanishing of the similar antiholomorphic current $j_-$. 
As in the case of the bosonic string $j_+$ and $j_-$ can be gauged away
by the residual holomorphic and antiholomorphic gauge transformations.
The holomorphicity follows from the equations of motion, see 
\cite{GrigorievTseytlinForthcoming} for a detailed discussion of these questions.
From the point of view of
the boost-invariant symplectic structure (\ref{SymplecticForm})
the equation $j_+=0$ can be interpreted as skew-orthogonality to the 
orbits of the gauge transformations $\delta g=g\xi$, $\delta K_1=[K_1,\xi]$,
$\xi\in \mathfrak{h}$.
Therefore $j_+$ can be understood as the moment map
of the ungauged mass-deformed WZW with fermions with respect to the symmetry:
\begin{eqnarray}
g(\tau^+,\tau^-) & \mapsto & g(\tau^+,\tau^-)h_R(\tau^+)\\
K_{1,3}(\tau^+,\tau^-) & \mapsto &
h_R(\tau^+)^{-1}K_{1,3}(\tau^+,\tau^-)h_R(\tau^+) \,.
\end{eqnarray}

\section*{Acknowledgments}

We thank A. Tseytlin for discussions and informing of and making available the 
forthcoming paper \cite{GrigorievTseytlinForthcoming}.
The research of AM was supported by the Sherman Fairchild 
Fellowship and in part
by the RFBR Grant No.  06-02-17383 and in part by the 
Russian Grant for the support of the scientific schools
NSh-8065.2006.2. 
The research of SSN was supported by a John A. McCone Postdoctoral Fellowship of Caltech. 

\appendix

\section{Hamiltonian reduction of WZW model}
\subsection{Classical string with periodic boundary conditions and 
Hamiltonian reduction of WZW}
\label{sec:HamiltonianReduction}
We have seen that on an infinite worldsheet the classical string is
equivalent to the gauged WZW model. The analysis on an infinite worldsheet is
formal because we neglect the boundary terms.
What happens when we consider instead a cylindrical worldsheet?
To understand periodic boundary conditions, we will use an interpretation
of the classical string as a Hamiltonian reduction. It turns out that
the classical string in the form described by Eqs. 
(\ref{gAAdata}) --- (\ref{LeftRightGaugeTransformations})
is closely related to  the Hamiltonian reduction of the mass deformed WZW model
with respect to the symmetries (\ref{SymmetriesToReducePrime}). 


\vspace{10pt}
\noindent
The precise relation is the following. Let ${\cal M}_{string}$ denote the space
of classical solutions of the equations 
(\ref{gAAdata}) --- (\ref{LeftRightGaugeTransformations}). It can be represented
as a continuous family of subspaces ${\cal M}_{string}^{[m_L],[m_R]}$
parametrized by the conjugacy classes of the monodromies
$m_L  =  \stackrel{\rightarrow}{P}
\exp \left[ -\int_0^{2\pi} \partial_-g g^{-1}|_{\mathfrak{h}}\; d\tau^- \right]$
and
$m_R  =  \stackrel{\rightarrow}{P}
\exp \left[ \int_0^{2\pi} g^{-1}\partial_+g|_{\mathfrak{h}}\; d\tau^+ \right]$:

\begin{equation}
{\cal M}_{string}=\bigcup_{[m_L],[m_R]} {\cal M}_{string}^{[m_L],[m_R]}\,.
\end{equation}
Each subspace ${\cal M}_{string}^{[m_L],[m_R]}$ is naturally identified
as the phase space of a 
Hamiltonian reduction of the mass deformed WZW model:
\begin{equation}
{\cal M}_{string}^{[m_L],[m_R]} = {\cal M}_{WZW/\!/(LH\times LH)}^{[m_L],[m_R]}
\end{equation}
where $[m_L]$ and $[m_R]$ are identified the conjugacy classes of the moment map.

\vspace{10pt}
\noindent We will explain in Section \ref{sec:HRvsGauged} that the Hamiltonian 
reduction of the
WZW model is closely related to the gauged WZW model. 

\subsection{Hamiltonian reduction of WZW model}
Consider a
classical mechanical system with the action of some group $H$ on the phase
space ${\cal M}$. Let $\mathfrak{h}=Lie(H)$ denote the Lie algebra of $H$.
Suppose that $H$ preserves the symplectic structure,
and therefore it is generated by a set of Hamiltonians;
each vector field $\xi\in \mathfrak{h}$ is generated by its own 
corresponding Hamiltonian ${\cal H}_{\xi}$. Notice that we should have
\begin{equation}\label{CentralExtension}
\{ {\cal H}_{\xi_1} , {\cal H}_{\xi_2} \} = 
{\cal H}_{[\xi_1,\xi_2]}+\mbox{const} \,.
\end{equation}
We think of the constant term as $H_{c}$ where $c$ is the central element
of some central extension of $\mathfrak{h}$, let us call it $\hat{\mathfrak{h}}$.
The {\em moment map} $\mu$ is a map from the phase space ${\cal M}$ to 
$(\hat{\mathfrak{h}})^*$,
which is defined in the following way. For each point $x\in {\cal M}$, 
and $\xi\in \mathfrak{h}$, we define:
\begin{equation}
\mu(x)\in \hat{\mathfrak{h}}^*\;: \;\;\;\langle\mu(x),\xi\rangle = H_{\xi}(x)
\end{equation}
It follows from (\ref{CentralExtension}) that the moment map has the property
of {\em equivariance}:
\begin{equation}
\mu(h.x)=\mbox{Ad}(h^{-1})^*.\mu(x)\,.
\end{equation}
The {\em Hamiltonian reduction} consists of three steps. First choosing a coadjoint 
orbit ${\cal O} \subset \mathfrak{h}^*$, then restricting to the subspace of the
phase space determined by the equation $\mu(x)\in {\cal O}$, and finally identifying
the points which are connected by the action of $H$: 
\[
x\simeq y \;\; \mbox{if}\;\; y=h.x\;\; \mbox{for some}\;\; h\in H \,.
\]
Schematically:
\begin{equation}
{\cal M}/\!/H = \mu^{-1}{\cal O}/H \,.
\end{equation}
Notice that  the Hamiltonian reduction depends on the choice of a coadjoint
orbit ${\cal O}\subset \mathfrak{h}^*$.

\vspace{20pt}
\noindent Let us now look at the Hamiltonian reduction of the WZW model by the symmetries:
\begin{equation}\label{SymmetriesToReducePrime}
g(\tau^+,\tau^-)\mapsto h_L(\tau^-) g(\tau^+,\tau^-) h_R(\tau^+)^{-1}
\end{equation}
where both $h_L$ and $h_R$ are in $H\subset G$.
The symplectic structure of the (ungauged) WZW model is given by this equation:
\begin{equation}\label{OmegaWZW}
\Omega_{WZW}=\int_0^{2\pi} d\tau^+\;\mbox{tr}\;
\delta g g^{-1}\partial_+(\delta g g^{-1}) \; - \;
\int_0^{2\pi} d\tau^-\;\mbox{tr}\;
g^{-1} \delta g\partial_-(g^{-1} \delta g) \,.
\end{equation}
Notice that these symmetries form two copies of the loop group\footnote{
The elements of the loop group $LH$ are group-valued functions $h(\sigma)$
satisfying $h(\sigma+2\pi)=h(\sigma)$}
 of $H$; therefore
the symmetry group is $LH\times LH$.
The infinitesimal version of (\ref{SymmetriesToReducePrime}) is
\begin{equation}
\delta_{(\alpha_L,\alpha_R)}g(\tau^+,\tau^-)=\alpha_L(\tau^-) g - 
g \alpha_R(\tau^+) \,,
\end{equation}
where the Lie algebra $L\mathfrak{h}\oplus L\mathfrak{h}$ is parametrized
by the pair $(\alpha_L,\alpha_R)$.
The moment map is:
\begin{equation}
\langle \mu \;,\; (\alpha_L,\alpha_R)\rangle =
-2\int_0^{2\pi} d\tau^- \mbox{tr}\;\alpha_L\partial_-g g^{-1}
-2\int_0^{2\pi} d\tau^+ \mbox{tr}\;\alpha_Rg^{-1} \partial_+g \,.
\end{equation}
This, in particular, implies that $\partial_+(\partial_-g g^{-1}|_{\mathfrak{h}})=
\partial_-(g^{-1}\partial_+g|_{\mathfrak{h}})=0$. 
We will denote:
\begin{equation}
j_+=g^{-1}\partial_+g|_{\mathfrak{h}} \qquad
\mbox{and} \qquad j_-=\partial_-g g^{-1}|_{\mathfrak{h}}
\end{equation}
The coadjoint action of 
$\widehat{\mathfrak{h}}\oplus \widehat{\mathfrak{h}}$ on $j_+$ and $j_-$ is given
by the formulas:
\begin{equation}\label{GaugeTransformationsAndEquivariance}
\begin{array}{lcl}
\delta_{\alpha_R} j_+ = -\partial_+\alpha_R - [j_+,\alpha_R] &\;\; &
\delta_{\alpha_R} j_- = 0 
\\
\delta_{\alpha_L} j_+ = 0&\;\; &
\delta_{\alpha_L} j_- = -\partial_-\alpha_L - [j_-,\alpha_L] \,.
\end{array}
\end{equation}
Since we want to discuss the Hamiltonian reduction of the WZW model, we need to know
what are the orbits of this coadjoint action.
To describe the orbits we need to classify the invariants of this action.
The invariants are the eigenvalues of the left and right monodromy matrices.
These monodromy matrices are defined as follows:
\begin{eqnarray}
m_L & = & \stackrel{\rightarrow}{P}\exp \left[ -\int_0^{2\pi} j_- d\tau^- \right] 
\label{DefM_L}
\\[5pt]
m_R & = & \stackrel{\rightarrow}{P}\exp \left[ \int_0^{2\pi} j_+ d\tau^+ \right] \,.
\label{DefM_R}
\end{eqnarray}

\subsection{How the space of solutions to  
(\ref{gAAdata}) --- (\ref{LeftRightGaugeTransformations})
is related to the Hamiltonian reduction of WZW by the symmetries
(\ref{SymmetriesToReducePrime})?} 
Let us look at the solutions to the system of equations 
(\ref{gAAdata}) --- (\ref{LeftRightGaugeTransformations}).
We denote this space ${\cal M}_{string}$.
Just as we did on the infinite line, we can still gauge away $A_+$ and
$A_-$ on the cylinder using the gauge transformations 
(\ref{LeftRightGaugeTransformations}); there is no obstacle.
In the gauge $A_{\pm}=0$
the equation (\ref{MassiveZeroCurvature}) becomes the WZW equation of motion
and the residual gauge transformations are precisely the
symmetries (\ref{SymmetriesToReducePrime}) which we used to define
the Hamiltonian reduction. 
On an infinite line we could use these residual gauge transformations
to put $j_{\pm}=0$, but on a cylinder the conjugacy classes of
$m_L$ and $m_R$ (defined in Eqs. (\ref{DefM_L}) and (\ref{DefM_R}))
are obstacles to gauging away $j_{\pm}$. The space of solutions
splits into a union of subspaces with a fixed conjugacy classes 
of $m_L$ and $m_R$:
\begin{equation}
{\cal M}_{string}=\bigcup_{([m_L],[m_R])\in H/H \times  H/H}
{\cal M}_{string}^{[m_L],[m_R]} \,.
\end{equation}
For every fixed $[m_L]$ and $[m_R]$ the subspace ${\cal M}_{string}^{[m_L],[m_R]}$
is identical to the phase space of the Hamiltonian reduction of the WZW model on the
value of the moment map corresponding to $([m_L],[m_R])$.

\subsection{How the Hamiltonian reduction of WZW is related to the
 gauged WZW?}\label{sec:HRvsGauged}
We want to explain in which sense the action of the massive gauged WZW
given by Eq. (\ref{ActionBPS}) describes
the classical string. We already explained how the classical string 
is related to the Hamiltonian reduction of the WZW model by the infinite
dimensional symmetry group $LH\times LH$ acting as
specified in (\ref{SymmetriesToReducePrime}).
But what is the relation between the Hamiltonian reduction of the WZW model
and the gauged WZW model? It turns out that these two models are
equivalent modulo subtleties with zero modes, which we will now 
describe.

We want to understand the 
relation between these two systems:
\begin{enumerate}
\item	Gauged WZW model defined by the action $S_{WZW}+S_{gauge}$
	(see Eq. (\ref{TermsInWZWAction}))
\item	Hamiltonian reduction of (ungauged) $S_{WZW}$ with respect to the symmetries:
\begin{equation}\label{SymmetriesToReduce}
g(\tau^+,\tau^-)\mapsto h_L(\tau^-) g(\tau^+,\tau^-) h_R(\tau^+)^{-1}
\end{equation}
\end{enumerate}
 As we explained, the procedure
of Hamiltonian reduction depends on the choice of a conjugacy class of $m_L$
and the choice of a conjugacy class of $m_R$.
In particular if $m_L=m_R={\bf 1}$, then we can use the gauge transformations 
(\ref{GaugeTransformationsAndEquivariance}) to choose $g$ to satisfy
(\ref{HolCurrentMustVanish}) and (\ref{AntiHolCurrentMustVanish}).
From this point of view Eq. (\ref{HolCurrentMustVanish})
defines a submanifold in the phase space which is skew-orthogonal with respect
to the Kirillov form (\ref{OmegaWZW}) 
to the orbit of the gauge transformations $g \mapsto gh_R^{-1}$,
see Section 5.6 of \cite{Mikhailov:2006uc}. Similarly (\ref{AntiHolCurrentMustVanish})
defines a subspace orthogonal to the orbit of $g\mapsto h_Lg$.

More generally, suppose that the conjugacy class of $j_+$ under
the transformation $\delta_{\alpha_R}$ coincides with the conjugacy class
of $j_-$ under the transformation $\delta_{\alpha_L}$.
This means that there is $f\in H$ such that
\begin{equation}\label{ConjugateMonodromies}
fm_Lf^{-1}=m_R\,.
\end{equation}
In this case we will denote
\[
{\cal M}_{string}^{[m]}={\cal M}_{string}^{[m],[m]} \,.
\]
It turns out that ${\cal M}_{string}^{[m]}$ can be identified as 
the Hamiltonian reduction of the gauged WZW phase space on the
level set of the conjugacy class of the holonomy 
of the WZW gauge field $A$ (the $A$ of Eq. (\ref{TermsInWZWAction})).

Indeed, let us describe the map from $g$ of (\ref{OmegaWZW}) to a solution
of the gauged WZW model. First of all, making the constant gauge
transformation with $f\in H$ we can choose 
\begin{equation}\label{MLEqualsMR}
m_L=m_R=m \,,
\end{equation}
and we can also rotate $m$ into a fixed maximal torus of $H$.
 Then consider $\widehat{g}$ defined
by the formula:
\begin{equation}
\widehat{g}(\tau,\sigma)=
\left(\stackrel{\rightarrow}{P}\exp \left[ -\int_0^{\sigma} j_- d\tau^- \right]\right)
g
\left(\stackrel{\leftarrow}{P} \exp \left[ -\int_0^{\sigma} j_+ d\tau^+ \right]\right) \,.
\end{equation}
Notice that $\widehat{g}$ has the following properties:
\begin{eqnarray}
&& \partial_-(\widehat{g}^{-1}\partial_+\widehat{g}) = 0
\\
&& \widehat{g}^{-1}\partial_+\widehat{g}|_{\mathfrak{h}}=
\partial_-\widehat{g}\widehat{g}^{-1}|_{\mathfrak{h}}=0
\\
&& \widehat{g}(2\pi)=m\widehat{g}(0)m^{-1} \,.
\label{PeriodicUpToConjugation}
\end{eqnarray}
This is almost what we need, except for we want to make $g$ periodic. 
Consider $\mu\in{\mathfrak h}$ such that
$m=e^{2\pi \mu}$, and define $\tilde{g}$:
\begin{equation}
\tilde{g}(\tau,\sigma) = e^{-\sigma \mu} \widehat{g}(\tau,\sigma) e^{\sigma \mu}
\end{equation}
One can check that $\tilde{g}$ satisfies the equations of motion
(\ref{GWZWEqM}), (\ref{AplusEqM}) and (\ref{AminusEqM}) of the gauged WZW model
with $A_{\tau}=0$ and $A_{\sigma}=\mu$. 

On the other hand, notice that any
solution of the gauged WZW can be gauged to $A_{\tau}=0, \; A_{\sigma}=\mu$
for some $\mu$. The conjugacy class $[\mu]$ is a dynamical variable
in the gauged WZW. But when we do the Hamiltonian reduction of the WZW model, we fix $[\mu]$.
Moreover, the Hamiltonian reduction of the WZW model has additional residual
 gauge tranformations
which correspond to the following transformations of $\tilde{g}$:
\begin{equation}\label{AdditionalResidualGaugeTransformations}
\delta \tilde{g}= \alpha \tilde{g} + \tilde{g}\alpha \quad , 
\qquad \alpha\in {\mathfrak h}
\end{equation}
where the gauge parameter $\alpha$ should commute with $\mu$: $[\alpha,\mu]=0$.
From the point of view of the gauged WZW model these transformations
are generated by the eigenvalues of $P\exp\int A$, {\it i.e.} the eigenvalues
of $\mu$.

\vspace{10pt}
{\small
\noindent Indeed, 
the symplectic form following from the action $S_{WZW}(g) + S_{gauge} (g, A_+, A_- )$
is:
\begin{eqnarray}
\Omega & = & \phantom{+}\int d\tau^+\; \mbox{tr}\;
\left( \delta g g^{-1} \nabla_+(\delta g g^{-1}) + 2\delta g g^{-1}\delta A_+\right)+
\nonumber
\\
&& + \int d\tau^-\; \mbox{tr}\;
\left(-g^{-1} \delta g \nabla_-(g^{-1} \delta g) + 2g^{-1} \delta g\delta A_-\right) \,.
\end{eqnarray}
We can choose the gauge where $A_+=-A_-=\mu$ and $\mu$ belongs to the Cartan
subalgebra of $\mathfrak h$. In this gauge it is straightforward to see that
the Hamiltonian $\mbox{tr}(\alpha \mu)$ generates 
(\ref{AdditionalResidualGaugeTransformations}).
}

\begin{figure}[th]
\centerline{\includegraphics[width=10cm]{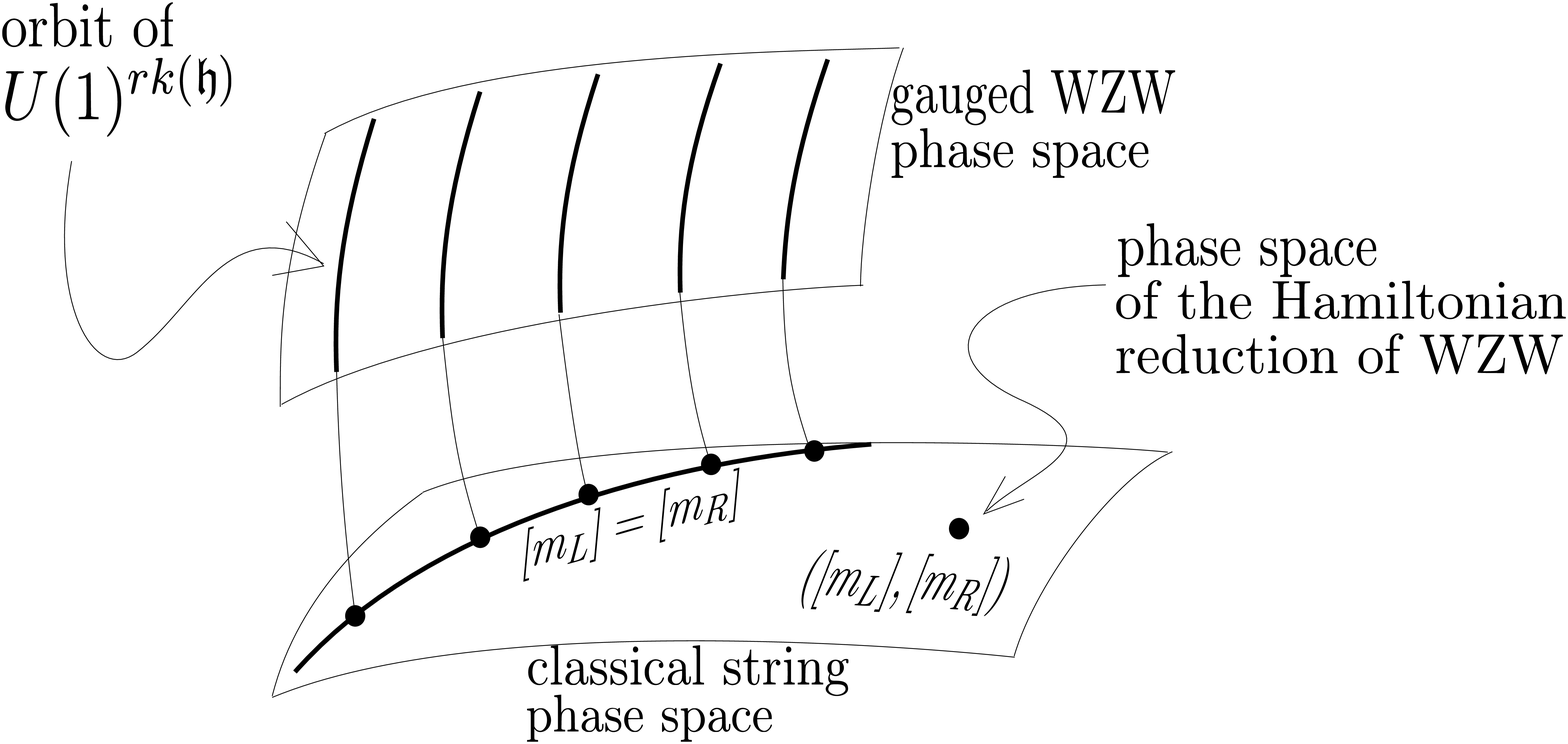}}
\caption{\label{fig:PhaseSpaces}\small 
The classical string phase space is shown as a horizontal plane;
each point on the plane corresponds to the phase space of the Hamiltonian reduction
of WZW for the value of the moment map $([m_L],[m_R])$. The phase space of the
gauged WZW is mapped on the subspace $[m_L]=[m_R]$ of codimension $rk(\mathfrak{h})$. 
The map involves identification
of the points related by the action of $U(1)^{rk(\mathfrak{h})}$.}
\end{figure}

\vspace{10pt}

\noindent We demonstrated that the Hamiltonian reduction of the WZW model on the 
fixed value $\mu$ of the moment map
corresponds to the gauged WZW with the fixed $P\exp\int A=e^{2\pi\mu}$ 
with the following identification: two configurations are considered equivalent
when they are related by the transformation 
(\ref{AdditionalResidualGaugeTransformations}). But this is precisely the
Hamiltonian reduction of the gauged WZW model on a fixed value
of the conjugacy class of the holonomy $P\exp\int A$.

The conclusion is that the Hamiltonian reduction of the WZW model
by the infinite-dimensional group $LH\times LH$ acting according to
Eq. (\ref{SymmetriesToReduce}) is equivalent to the Hamiltonian
reduction of the gauged WZW model by the finite-dimensional group
$U(1)^{rk(\mathfrak{h})}$ generated by the eigenvalues of the 
holonomy of the gauge field.

Figure \ref{fig:PhaseSpaces} illustrates the relation between the
phase spaces.


\bibliographystyle{JHEP}
\renewcommand{\refname}{Bibliography}
\addcontentsline{toc}{section}{Bibliography}


\providecommand{\href}[2]{#2}\begingroup\raggedright
\endgroup


\end{document}